\documentclass[conference, 10pt]{IEEEtran}
\IEEEoverridecommandlockouts
\pdfoutput=1 
\usepackage{cite}
\usepackage{amsmath,amssymb,amsfonts}
\usepackage{algorithmic}
\usepackage{graphicx}
\usepackage{textcomp}
\usepackage{xcolor}
\usepackage{tikz}
\usetikzlibrary{shapes,decorations}
\usepackage{amsthm}
\usetikzlibrary{automata, positioning}
\usepackage{relsize}

\usepackage{subfig}
\usepackage[font=footnotesize]{caption}

\usepackage{multicol}
\usepackage{bbm} 
\usepackage{dsfont}
\usepackage{macros}
\tikzset{fontscale/.style = {font=\relsize{#1}} }

\def\BibTeX{{\rm B\kern-.05em{\sc i\kern-.025em b}\kern-.08em
		T\kern-.1667em\lower.7ex\hbox{E}\kern-.125emX}}
\begin{document}
	
	\title{Cacheable and Non-Cacheable Traffic Interplay in a Relay-Assisted Wireless Network}
	
	\author{
		\IEEEauthorblockN{Ioannis Avgouleas, Nikolaos Pappas, and Vangelis Angelakis} 
		\IEEEauthorblockA{Department of Science and Technology, Link\"oping University \\
									Norrk\"oping, 60174, Sweden }
		E-mails: \{ioannis.avgouleas, nikolaos.pappas, vangelis.angelakis\}@liu.se 
	}

\thispagestyle{plain}
\pagestyle{plain}

\IEEEpeerreviewmaketitle 

\maketitle
	
\begin{abstract}
We study a discrete-time wireless network that serves both cacheable and non-cacheable traffic with assistance of a relay node with storage capabilities for both types of traffic. 
We investigate how allocating the storage capacity to cacheable and non-cacheable traffic affects the network throughput.
Our numerical results provide useful insights by varying not only the allocation of cacheable to non-cacheable storage but also the rate by which non-cacheable content is transmitted, the rate by which cacheable content is requested, as well as different popularity distributions of the cached files.
\end{abstract}

\section{Introduction}
Wireless data traffic grew immensely over the past 10 years and is expected to comprise 20 percent of total IP traffic by 2022 with 1.5 mobile-connected device per capita. 
Almost three-fifths of traffic will be offloaded from cellular networks to Wi-Fi by the same year \cite{Cisco_Data}. 
Mobile devices represented a large part of the total wireless traffic with wireless video being one of the main sources of wireless data traffic. 
Moreover, the introduction of high quality video formats such as 4K, 8K, 360$^o$ etc. will further contribute to degraded user experience due to increased delay and congestion.
Video streaming services are interested in mitigating such performance issues by offloading content closer to the users 
\cite{cache_for_streaming_video}.

Additionally, the advent of the Internet of Things (IoT) will necessitate serving a massive amount of devices with limited resources of energy, memory and computation. 
Thus, the attention of the research community is moving towards effectively supporting IoT communications.
For instance, Named Data Networking (NDN) is an information-centric Internet architecture which has been recently considered as an enabling technology for IoT, due to its innovative features like named-based routing and in-network caching \cite{NDN}.
NDN allows for caching at intermediate nodes which proves to be effective in mitigating the network delay and traffic as well as the load on content producers \cite{Caching_NDN}.

Caching has been quite successful recently in reducing cellular traffic and delay as well as increasing network throughput and reliability.
Cache-enabled 5G wireless systems and future network architectures will benefit from caching in terms of reduced costs for the network infrastructure and the quality of service available to the end users.
A cache can typically store a small subset of the files library because of its limited capacity. 
Therefore, caching policies are necessary to decide which files are placed into the cache as well as which files to evict from the cache when the cache is full and a new file should be cached.  
Many content placement strategies have been proposed in the literature e.g., caching the least frequently used content \cite{Paschos_IEEE_Comm_Mag}, caching the most popular content everywhere \cite{Paschos_IEEE_Comm_Mag}, probabilistic caching \cite{ProbabilisticCaching}, cooperative caching \cite{CooperativeCaching}, and geographical caching \cite{GeographicalCaching}. 
The role of caching for future communication systems is analyzed in \cite{Caching_Survey}.

Caching policies usually assume a model for file requests which, in practice, is a priori unknown and time-varying \cite{Ephemeral:Content:Popularity}. 
Additionally, there are caching architectures whose caches receive a low number of requests and, thus, realize request processes with high non-stationary popularity \cite{NonStationaryRequests1}.
However, such models are challenging to fit and depend on strong assumptions about the popularity distribution.
The authors in \cite{Paschos_Learning2Cache} develop a class of policies that make no assumptions on the file request distribution and, hence, is robust to popularity deviations by adjusting their caching decisions when the popularity model changes. 

Typical performance criteria for a caching policy include the cache hit ratio (or probability), the density of successful receptions \cite{ProbabilisticCaching}, energy efficiency\cite{EnergyEfficiency} and the traffic load of the wireless links \cite{TrafficLoad}, among others.
Additionally, a considerable amount of contemporary works consider throughput and/or delay with caching helpers \cite{Caching:Helpers, Caching:Helpers:Multiple,Caching:Secrecy}. 
To reduce delay, many works mitigate the backhaul or transmission delay under the assumption that traffic or requests are saturated. Works that do not make this assumption, but assume stochastic arrivals have also appeared e.g., \cite{CachingNetsAnalysis}.

The rise of wireless networks serving a massive amount of devices, such as 5G or IoT networks, will give birth to new traffic patterns.
For example, traffic generated from Machine to Machine (M2M) devices is generally different compared to traditional smartphone traffic \cite{M2M:traffic, Traffic:Classification}. 
These findings along with the aforementioned proliferation of caching techniques at the network edge suggest that understanding traffic is key to designing and optimizing the performance of future networking architectures.

\subsection{Our work}
In this paper, we study a wireless system that serves both cacheable and non-cacheable traffic. 
A relay node partially assists the non-cacheable transmissions by queueing the non-cacheable packets that failed to be transmitted to the destination. 
The queued packets are intended to be transmitted from the relay to the destination in a subsequent time slot.
\begin{figure}[t!]	
	\centering
	\includegraphics[width=1\linewidth]{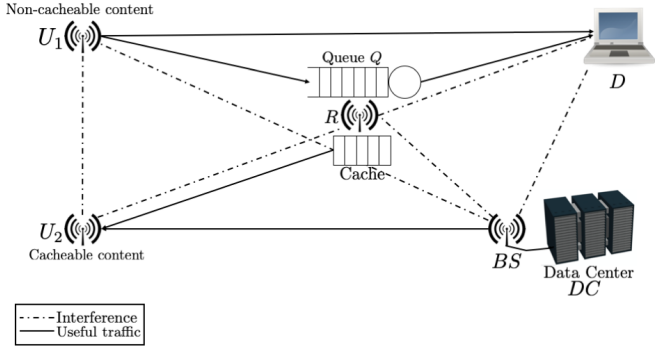}
	\caption{An example configuration of our system model. User $U_1$ sends non-cacheable traffic to the destination user $D$ with the assistance of relay node $R$. The storage capabilities of $R$ can be split between cacheable and non-cacheable traffic. 
	In case $U_1$'s transmissions to $D$ fails, the failed packet is stored at the relay's queue $Q$ given that there is a successful transmission from $U_1$ to $R$. 
	User $U_2$ has a cache and requests cached files from external resources with some probability. The relay node can serve $U_2$ given that it has the requested file and it is not serving $U_1$. Otherwise, the requested cached file can be fetched from the data center $DC$ through the base station $BS$. The data center hosts the entire library of files and is available to serve the requests cached files with some probability. 
	}
	\label{Fig:SystemModel}		
\end{figure}
Moreover, when the relay node is not assisting the non-cacheable pair, it is available to serve cached files to another wireless user within in its coverage.
The wireless user that requests cached content can also be served by a data center in case the relay misses the requested file or is not available for caching. 
The data center is assumed to contain the file library and is connected directly to a wireless base station through a backhaul link. Files from the data center are fetched to the cached user through the base station if the data center is available to serve cached files.

We analyze the network throughput considering the rate by which non-cacheable traffic is transmitted to the destination as well as the rate by which the relay attempts transmissions for non-cacheable traffic.
In our numerical results, we vary the storage capacity dedicated to non-cacheable traffic and, thus, to cacheable traffic, to gain useful insights into the throughput of such systems and introduce the first step for the understanding of larger network topologies.

\section{System Model}\label{Sec:SystemModel}
\subsection{Network Model}
We consider the following network system: a user device $U_1$ serving non-cacheable traffic to a destination node $D$ with the assistance of a wireless relay node $R$, and another user, $U_2$, requesting cached files in case of a local cache miss. The requested cached file can be served by external resources i.e., from the relay node $R$ or the data center $DC$. We assume that packets and files are equally sized, so, thereafter, we use the terms packet and files interchangeably. 
The topology of the studied system can be found in Fig. \ref{Fig:SystemModel}.

We assume slotted time and that a packet transmission takes exactly one time-slot. 
Nodes have random access to the wireless medium with no coordination between them regarding transmissions' scheduling. Thus, nodes attempt transmissions to the channel with some probability.
An acknowledgment mechanism is assumed such that instantaneous and error-free acknowledgment/ negative-acknowledgment (ACK/NACK) packets are sent by the receiver over a separate channel.
As a result, when $D$ successfully receives a packet from $U_1$, the latter removes it from its buffer and is ready to attempt transmission of the next packet (in the next time slot) and the relay $R$ discards it from its queue (if it has successfully received it). 
When $R$ successfully receives a packet that did not reach $D$, node $U_1$ discards it from its buffer and is ready to attempt transmission of the next packet.
The evolution of the relay's queue is analysed in Section \ref{Sec:QueueAnalysis}.

Additionally, the relay node $R$ does not generate packets on its own and is equipped with a FD transceiver i.e., it can receive and transmit a packet within the same time slot. 
Its purpose is two-fold: (i) forward non-cacheable packets to the destination node, and (ii) serve cached files to user $U_2$. In each time slot, the relay is available to serve either non-cacheable or cacheable traffic.
For that purpose, it hosts $F$ files that can be used for serving both types of traffic. 
Non-cacheable incoming packets to $R$ are stored in its queue with size for $B$ packets. The rest of the storage capacity is devoted to cached files for user $U_2$.
Consequently, the cache at the relay can hold $F-B$ files to help user $U_2$'s file requests from external resources. The operation of the relay's cache is described in Section \ref{Sec:CachePolicy}.

The data center can be accessed through a wireless base station ($BS$) and stores the library of files i.e., all files that $U_2$ might request. 
We model $DC$'s availability with \pro $\alpha$ to model the fact that it can be out of service due to serving other users, failure, maintenance etc. If the $DC$ is always available to $U_2$, then $\alpha = 1$. On the other hand, if the $DC$ is not available for $U$, then $\alpha = 0$.

\subsection{Transmission Model}
\begin{table}[bp] \caption{Notation table}
	\def\arraystretch{1.1}
	\begin{center}
		\begin{tabular}{ | c  l |}
			\hline
			\textbf{Notation} & \textbf{Description} \\ \hline
			$q_{1}$   & probability of $U_1$ attempting non-cacheable transmission.\\
			$q_{R}$   & probability of $R$ being available to serve $D$.\\
			$q_{U}$   & probability of $U_2$ requesting a cached file from external\\
			&  resources ($R$ or $BS$).\\			
			$p_{h}$ & probability of cache hit at $R$. \\ 
			$\alpha$ & probability of $DC$ being available to serve $U_2$ requests.\\ 
			\hline
			$\PSNR{i}{j} $ & success probability of link $i \rightarrow j$, when node $i$ is the \\
			& only transmitter.\\ 
			$\PSINR{i}{j}{T} $ & success probability of link $i \rightarrow j$, when $i$ and nodes\\
			& in $T$ are transmitting.\\ 
			\hline  
		\end{tabular}
	\end{center}
	\label{table:probabilities}
\end{table}
In each time slot, $U_1$ attempts to transmit non-cacheable traffic to $D$ with probability $q_1$ and the relay $R$ can serve non-cacheable traffic to $D$ with probability $q_R$.
Thus, it is available to serve cacheable traffic to $U_2$ with \pro $1-q_R$.
Moreover, in each time slot, $U_2$ requests a cached file from external resources with probability $q_U$. 
Therefore, if the relay node serves $D$ with non-cacheable traffic, then $R$ interferes with the transmission from $U_1$ to $D$ and any transmission from $BS$ to $U_2$ which happens when $U_2$ has requested a file which will be served by the data center $DC$.
On the other hand, if the relay node serves $U_2$ with cacheable traffic, then it only interferes with the transmission from $U_1$ to $D$.
We summarize the aforementioned events and notation in Table \ref{table:probabilities}. 

\subsection{Physical Layer Model}\label{Sec:LinkSuccessProbabilities}
We assume Rayleigh fading for the wireless channel and that a packet transmission from node $i$ to node $j$ is successful if and only if the link Signal-to-Interference-plus-Noise power ratio (SINR) between node $i$ and $j$ exceeds a minimal threshold $\theta$. 
The received power at node $j$ when $i$ transmits is $P_{rx}(i,j) = A(i,j)h(i,j)$, where $A(i,j)$ is a unit-mean exponentially distributed random variable and the received power factor is:  
$ h(i,j) = P_{tx}(i)/ r(i,j)^p,$
where $P_{tx}(i)$ is the power measured at $1~m$ away from the transmitting antenna of node $i$, $r(i,j)\geq1~m$ is the distance in $m$ between $i$ and $j$, and $p$ is the path-loss exponent.

\noindent The success probability of link $i \rightarrow j$, with $\mathcal{T}$ denoting the set of transmitting nodes, is given by \cite{Rayleigh:fading:success:prob}:
\begin{align*}
\PSINR{i}{j}{\mathcal{T}} &= exp\Big(-\theta \frac{ n_j}{h(i,j)}\Big)
\prod_{k \in \mathcal{T} \setminus \{ i,j \}} \Big( 1 + \theta\frac{ h(k,j)}{h(i,j)} \Big)^{-1},
\end{align*}
where $n_j $ is the noise power at receiver $j$.

\subsection{Caches' Operation}\label{Sec:CachePolicy}
We assume the content placement is given and that the cached nodes i.e., $U_2$ and $R$, follow the Collaborative Most Popular Content (CMPC) policy. According to the latter, user $U_2$ stores the first $M_U$ most popular files in its cache, the relay node caches the next $F-B$ most popular files, and the data center hosts all files that $U_2$ might request.
When the user node requests for a file that is not stored in its most popular files, it first probes $R$ for it. The relay node serves $U_2$ if it is available for caching i.e., when not serving $U_1$, and has the requested file.
Otherwise, user $U_2$ requests the file from the data center. 
If the latter is available for $U_2$ (which happens with probability $\alpha$), then the file is fetched by the data center.
We assume that the information exchange required for the operation of the CMPC policy e.g., the cache size of each device and the content placement in each device, is negligible.

Moreover, we consider a finite content library of $N$ files with $f_i$ denoting the $i$-th most popular file.
For the sake of simplicity, we assume that all files have equal size and that access to cached files happens instantaneously.
The request probability of the $f_i$ is given by: $p_i = \Omega / i^\delta$, where $\Omega = ( \sum_{j=1}^{N}  j^{-\delta}   )^{-1}$ is the normalization factor and $\delta$ is the shape parameter of the Zip law which determines the correlation of user requests.
As a result, the probability that user $U_2$ requests a file that is not located in its cache is:
$ q_U  = 1 - \sum_{i=1}^{M_U} p_i, $
and the cache hit probability at the relay node $R$ is given by: 
$ p_h  = \sum_{i=M_U+1}^{M_U+F-B} p_i,$
where $F$ and $B$ are the storage capacity and the queue size at the relay node, respectively.
\section{Analysis}\label{Sec:Throughput}
In this section, we present the analysis for the states of the relays' queue and the throughput of the system in Fig. \ref{Fig:SystemModel}.
\subsection{Relay's Queue Analysis}\label{Sec:QueueAnalysis}
\begin{figure}[!h]
	\centering
	\includegraphics[width=1\linewidth]{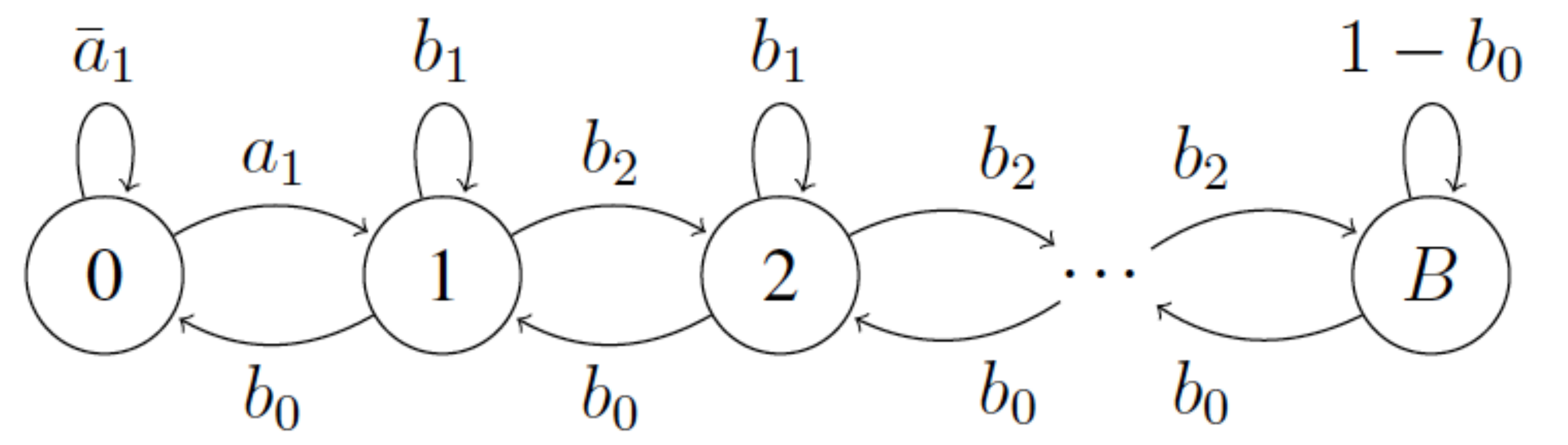}
%
%
%
	\caption{Markov Chain of the relay's queue.} \label{fig:DTMC}
\end{figure}

Let $B$ denote the buffer size of the relay's queue $Q$. The latter follows the first-come-first-serve (FCFS) discipline.
Moreover, we assume that when the queue is full i.e., holds $B$ packets, and a new packet arrives at the relay, the relay rejects its arrival and acknowledgments user $U_1$ to attempt retransmitting that packet in a subsequent time slot. We consider that this acknowledgment is instantaneous and error-free.
The evolution of the Discrete Time Markov Chain (DTMC) of $Q$ is shown in Fig. \ref{fig:DTMC}.

\noindent The transition matrix that models the DTMC above is given by the following stochastic column matrix:
\begin{equation*}
\textbf{P} = 
\begin{pmatrix}
\begin{matrix}
\bar{a}_{1} 	 & b_0 	   & 		    	& 	      &    	   \\
a_{1} 	 & b_1  	& \ddots	&  		 &    	  \\
& b_2 	  	& \ddots   & b_0 & 		   \\ 
& 				& \ddots   & b_1 &	b_0 \\
&  				&  				& b_2 &	\bar{b}_0 
\end{matrix}
\end{pmatrix},
\end{equation*}
where $\bar{q} = 1 - q$. The entries of $P$ are given by:
\begin{align*} 
a_1  
		& = ~ \mathbb{P}(\textit{``Q increases by 1 packet when Q=0"}) = \\
		&= ~ q_1 \bar{q}_U(1-\PSNR{1}{D}) \PSNR{1}{R} \\
		&+ ~ q_1 q_U p_h  (1-\PSINR{1}{D}{R}) \PSNR{1}{R} \\ 
		&+ ~ q_1 q_U \bar{p}_h \alpha (1-\PSINR{1}{D}{BS}) \PSINR{1}{R}{BS} \\
	 	&+ ~ q_1 q_U \bar{p}_h \bar{\alpha} (1-\PSNR{1}{D}) \PSNR{1}{R}, \\
b_0  
	 &=  ~ \mathbb{P}(\textit{``Q decreases by 1 packet when Q$>$0"}) = \\ 
	 &= ~ q_R \bar{q}_1 \big( \bar{q}_U \PSNR{R}{D}  + q_U \bar{p}_h (\alpha \PSINR{R}{D}{BS} +\bar{\alpha} \PSNR{R}{D}) \big)\\
	& + ~ q_R q_1 \bar{q}_U \times \\
	& ~ \big[ \PSINR{R}{D}{1} \big( \PSINR{1}{D}{R} + (1-\PSINR{1}{D}{R})(1-\PSNR{1}{R}) \big) \big] \\
	& + ~ q_R q_1 q_U \bar{p}_h \big(  \alpha \big[ \PSINR{R}{D}{1, BS} \big(\PSINR{1}{D}{R,BS}+ \\ 
	&  ~(1-\PSINR{1}{D}{R, BS}) (1-\PSINR{1}{R}{BS}) \big) \big]+ \\
	& ~ \bar{\alpha} \big[ \PSINR{R}{D}{1} \big( \PSINR{1}{D}{R} +  (1-\PSINR{1}{D}{R}) (1-\PSNR{1}{R})\big) \big] \big), \\ 
b_1 & = ~ 1 - b_0 - b_2 = \mathbb{P}(\textit{``Q does not change"}), \\ 
b_2 
 	   & =  ~\mathbb{P}(\textit{``Q increases by 1 packet when Q$>$0"}) = \\
 	   & = ~ q_1 q_R \bar{q}_U (  1-\PSINR{R}{D}{1} )  (  1-\PSINR{1}{D}{R} ) \PSNR{1}{R} \\ 
       & +~ q_1 q_R q_U \bar{p}_h \alpha (  1-\PSINR{R}{D}{1, BS} ) \times \\
 	  & (  1-\PSINR{1}{D}{R, BS} ) \PSINR{1}{R}{BS} \\
 \end{align*}
\begin{align*} 
 	  &	+ ~ q_1 q_R q_U \bar{p}_h \bar{\alpha} (  1-\PSINR{R}{D}{1} )  (  1-\PSINR{1}{D}{R} ) \PSNR{1}{R} \\
 	  &	+ ~ q_1 \bar{q}_R [ \bar{q}_U (1-\PSNR{1}{D}) \PSNR{1}{R} ~+\\
 	  &~ q_U p_h (1-\PSINR{1}{D}{R}) \PSNR{1}{R}] \\
 	 & + ~ q_1 \bar{q}_R q_U \bar{p}_h \alpha (1-\PSINR{1}{D}{BS}) \PSINR{1}{R}{BS} \\
 	 &	+ ~ q_1 \bar{q}_R q_U \bar{p}_h \bar{\alpha} (1-\PSNR{1}{D}) \PSNR{1}{R}, \\
\end{align*}
To derive the steady state distribution $ \pi = [\pi_0, \cdots, \pi_B]^T $, we need to solve the balance equations: 
$ \textbf{P} \pi = \pi $
which produce the following relation for calculating the probability of being in state $i$: 
\[  \pi_i = \rho^{i-1} t_0 \pi_0, ~\forall~1 \leq i \leq B, \text{ and }
\pi_0 = \Big[ 1 + t_0 \Big( \frac{1-\rho^{B}}{1-\rho}  \Big)  \Big]^{-1}, \]

where:
$  \rho = b_2 / b_0~\text{  and  }~t_0 = a_1 / b_0. $
%
%
%
%
%
\subsection{Throughput Analysis}
First, we derive the direct throughput from user $U_1$ to the destination node $D$ and the relayed throughput from the relay node $R$ to $D$ with the intention of deriving the throughput of $D$. 
Then, we formulate the cacheable throughput seen by user $U_2$.
Recall that user $U_1$ and the relay attempt transmissions with probabilities $q_1$ and $q_R$, respectively. Moreover, $U_2$ requests a file from external resources with probability $q_U$, and $p_h$ is the probability of a cache hit at the relay's cache. 
Also, $\alpha$ denotes the probability with which the data center is available to serve file requests of $U_2$ and $\pNotZero$ denotes the probability that $Q$ i.e., the queue at the relay, is not empty (please see Table \ref{table:probabilities} for the description of our notation).

\noindent The \textit{direct throughput from user $U_1$ to $D$} is:
\begin{align*} \nonumber
T_{1 \rightarrow D} = &~ q_1 q_R \pNotZero  \big[ q_U \bar{p_h} \alpha \PSINR{1}{D}{R, BS} \\
+&~ q_U \bar{p}_h \bar{\alpha} \PSINR{1}{D}{R} + \bar{q}_U \PSINR{1}{D}{R}  \big]  \\ 
+&~ q_1 [1 -q_R \pNotZero ]  \big[\bar{q}_U \PSNR{1}{D} + q_U p_h \PSINR{1}{D}{R} \\
+&~ q_U \bar{p}_h \alpha \PSINR{1}{D}{BS} + q_U \bar{p_h} \bar{\alpha} \PSNR{1}{D} \big]. \\
\end{align*}
The \textit{relayed throughput from $R$ to $D$} is given by: 
\begin{align*}
 T_R =& q_1 P(Q=0)  \big[ \bar{q}_U (1-\PSNR{1}{D})  \PSNR{1}{R} \\
 		+ & q_U  p_h (1-\PSINR{1}{D}{R}) \PSNR{1}{R}\\
 	 	+ & q_U \bar{p}_h \alpha (1-\PSINR{1}{D}{BS})\PSINR{1}{R}{BS} \\
 	 	+ & q_U \bar{p}_h \bar{\alpha} (1-\PSNR{1}{D}) \PSNR{1}{R} \big] \\
 	   	+ & q_1 P( 0<Q<B ) q_R \big[  \bar{q}_U (1-\PSINR{1}{D}{R}) \PSNR{1}{R} \\
 	   	+ & q_U \alpha (1-\PSINR{1}{D}{R,BS}) \PSINR{1}{R}{BS} \\
		+ & q_U \bar{\alpha} (1-\PSINR{1}{D}{R}) \PSNR{1}{R}  \big]\\
  	   	+ & q_1 P( 0<Q<B ) \bar{q_R} \times \\
		   & \big[  \bar{q}_U (1-\PSNR{1}{D}) \PSNR{1}{R} + \\
 		 & q_U p_h (1-\PSINR{1}{D}{R} ) \PSNR{1}{R} +  q_U \bar{p}_h \times \\
		&\big( \alpha (1-\PSINR{1}{D}{BS}) \PSINR{1}{R}{BS} + \bar{\alpha} (1-\PSNR{1}{D}) \PSNR{1}{R}\big)\big] \\
  	   	+ & q_1 P(Q=B) q_R  \big[ \bar{q}_U \PSINR{R}{D}{1} (1-\PSINR{1}{D}{R})\PSNR{1}{R} \\
   \end{align*}
\begin{align*}
  	   	+ & q_U \alpha \PSINR{R}{D}{1, BS} (1-\PSINR{1}{D}{R, BS}) \PSINR{1}{R}{BS} \\ 
  	    + & q_U \bar{\alpha}  \PSINR{R}{D}{1}(1-\PSINR{1}{D}{R})\PSNR{1}{R} \big].
\end{align*}
The \textit{non-cacheable throughput seen by $D$} is given by:
\[  T_D = T_{1 \rightarrow D} + T_R. \]

The \textit{cacheable throughput seen by $U_2$} is given by:
\begin{align*}
T_2 = &~ q_U q_R \pNotZero  \bar{p}_h\alpha [ q_1 \PSINR{BS}{2}{R,1}  +  \bar{q}_1 \PSINR{BS}{2}{R}]  \\
       + &~ q_U [1 - q_R \pNotZero]  q_1 	   [p_h \PSINR{R}{2}{1} + \bar{p}_h \alpha \PSINR{BS}{2}{1}] \\ 
       + &~ q_U [1 - q_R \pNotZero] \bar{q}_1  [p_h \PSNR{R}{2} + \bar{p}_h \alpha \PSNR{BS}{2}]. 
\label{eq:ThroughputAt2}
\end{align*}

%
%
%
%
%
%
%
%
\section{Numerical Results}\label{Sec:Results}
In this section, we present numerical evaluations of the analysis in Section \ref{Sec:Throughput}.
The transmission power of each device and the distances between nodes are set as per Table \ref{table:links_params}. 
Please notice that we use the same SINR $\theta$ for every wireless link in our system and the same noise power $n$ for every receiver.
\vspace{1mm}
\begin{table}[bhp] \caption{Wireless links parameters for our numerical results.} 
	\scriptsize
	\def\arraystretch{1.1}
	\centering
		\begin{tabular}{ | c  r |  } 
			\hline \textbf{Parameter} & \textbf{Value} \\ \hline
			\hline
			$P_1$ & $ 1 ~mW$ \\ 	
			$P_R $ & $2 ~mW$ \\ 
			$P_{DC}$ & $ 10 ~mW$ \\ 
			$ n $ & $ 10^{-11} ~W $ \\ 	
			$ p $ & $ 4 $ \\ 
			$ \theta $ & $0$ or $5$ dB \\ 
			\hline
	\end{tabular}
	\begin{tabular}{ | c  r |  } 
	\hline \textbf{Parameter} & \textbf{Value} \\ \hline
	\hline
			$ r(1, 2)$  & $ 100 ~m$ \\
			$ r(1, BS)$  & $ 100\sqrt{2} ~m $ \\
			$ r(1, D) $ & $ 100 ~m$ \\ 
			$ r(1, R)$  & $ 50\sqrt{2} ~m $ \\
			$ r(BS, 2)$ & $ 100 ~m $ \\
			$ r(D, BS)$ & $  100 ~m $ \\
			$ r(R, 2)$ & $ 50\sqrt{2} ~m $ \\ 
			$ r(R, BS)$ & $   50\sqrt{2} ~m $ \\  
			$ r(R, D)$ & $ 50\sqrt{2}~m $ \\ 
		\hline
		\end{tabular} \\
		\vspace{1mm}
				\begin{tabular}{ | c  c |  } 
					\hline \textbf{Parameter} & \textbf{Value} \\
					$\theta $ & $0~dB$ \\
					\hline
					$ P_{ 1 \rightarrow D} $	   	& $ 0.368   $ \\ 
					$ P_{ 1 \rightarrow D/R} $	   & $ 0.041  $ \\ 
					$ P_{ 1 \rightarrow D/BS} $	  & $  0.033 $ \\ 
					$ P_{ 1 \rightarrow D/R,BS}$ & $ 0.004 $ \\
					\hline
					$ P_{ 1 \rightarrow R }$   	    & $  0.779$ \\ 
					$ P_{ 1 \rightarrow R/BS }$   & $  0.071$ \\
					\hline
					$ P_{ R \rightarrow D }$  		  & $ 0.883 $ \\ 
					$ P_{ R \rightarrow D/1} $	    & $ 0.784 $ \\ 
					$ P_{ R \rightarrow D/BS} $	   & $  0.392$ \\ 
					$ P_{ R \rightarrow D/1,BS} $ & $  0.349$ \\ 			
					\hline
					$ P_{ BS \rightarrow 2 }$       & $  0.905$ \\ 
					$ P_{ BS \rightarrow 2/1} $	   & $  0.823$ \\ 
					$ P_{ BS \rightarrow 2/R} $	   & $  0.503$ \\ 
					$ P_{ BS \rightarrow 2/1,R}$  & $  0.457$ \\ 			
					\hline
					$ P_{ R \rightarrow 2 }$  		 & $ 0.883$ \\ 
					$ P_{ R \rightarrow 2/1} $	   & $ 0.784$ \\
					\hline
				\end{tabular} 
				\begin{tabular}{ | c  c |  } 
					\hline \textbf{Parameter} & \textbf{Value} \\
					$\theta $ & $5~dB$ \\
					\hline
					$ P_{ 1 \rightarrow D} $	   	& $  0.042  $ \\ 
					$ P_{ 1 \rightarrow D/R} $	   & $  0.002 $ \\ 
					$ P_{ 1 \rightarrow D/BS} $	  & $   0.001$ \\ 
					$ P_{ 1 \rightarrow D/R,BS}$ & $  0$ \\
					\hline
					$ P_{ 1 \rightarrow R }$   	    & $  0.454$ \\ 
					$ P_{ 1 \rightarrow R/BS }$   & $  0.014$ \\
					\hline
					$ P_{ R \rightarrow D }$  		 & $  0.674$ \\ 
					$ P_{ R \rightarrow D/1} $	    & $  0.483$ \\ 
					$ P_{ R \rightarrow D/BS} $	   & $  0.136$ \\ 
					$ P_{ R \rightarrow D/1,BS} $ & $  0.098$ \\ 			
					\hline
					$ P_{ BS \rightarrow 2 }$        & $  0.729$ \\ 
					$ P_{ BS \rightarrow 2/1} $	   & $  0.554$ \\ 
					$ P_{ BS \rightarrow 2/R} $	   & $  0.207$ \\ 
					$ P_{ BS \rightarrow 2/1,R}$  & $  0.157$ \\ 			
					\hline
					$ P_{ R \rightarrow 2 }$  		 & $  0.674$ \\ 
					$ P_{ R \rightarrow 2/1} $	   & $   0.483$ \\
					\hline
				\end{tabular} 	
	\label{table:links_params}
\end{table}
Regarding caching, we assume that the cache of node $U_2$ hosts $M_U$ files, and that the relay node stores $F=10$ files for both types of traffic. Its queue has finite size $B$ for cacheable traffic and, hence,  $R$ holds $F-B$ files the non-cacheable traffic. 
We also assume that the whole library (at the data center) holds $N = 10000$ files.
The caches follow the Collaborative Most Popular Content (CMPC) policy (which we describe in Section \ref{Sec:CachePolicy}). The random availability of the data center for $U_2$ was set to $\alpha = 0.7$. We summarize the cache parameters in Table \ref{table:caches_params}.
In the following results, we vary $B \in [0,F]$ to gain insight into its effect on the throughput at the destination nodes ($U_2$ and $D$) and the distribution of the packets at the queue of the relay $R$.

\begin{table}[!b] \caption{Cache parameters and attempt probabilities for the transmitters in our numerical results.}
	\scriptsize
	\def\arraystretch{1}
	\centering
	\begin{tabular}{ | c l c |} 
		\hline \textbf{Parameter} & \textbf{Description}  & \textbf{Value} \\ \hline
		\hline
		$ M_U $   & cache size at user $U_2$			  										   & $0$ or $5$   \\ 
		$ F $		 & number of files  at the relay $R$    	   									&   $10$   \\ 
		$ N     $   & number of files at the data center $DC$			 					 & $10000$ \\ 
		$ \delta$  & shape parameter for the correlation of the files 	   & $0.5$ or $1.2$\\ \hline
		$ \qu $  	& probability of $U_2$ requesting a cached file    		&  $ 0.984$ or $1$ \\
		$q_R$   & probability of $R$ being available to serve $D$.       & $0.8$ \\
		& 	from external resources ($R$ or $BS$)													 &  \\
		$ \alpha $ & probability of $DC$ being available to serve $U_2$   & $ 0.7 $  \\ 
		$ \theta $  & links SINR minimum value				 						  & $ 5 $ dB  \\ 
		\hline
	\end{tabular}
	\label{table:caches_params}
\end{table}
\subsection{Throughput $T_D, T_2,$ and $T$ vs. Queue Size $B$}\label{Sec:Throughputs}
\begin{figure}[t]
	\centering
	\subfloat[$q_1 = 0.4$.]{
		\includegraphics[width=0.7\linewidth]{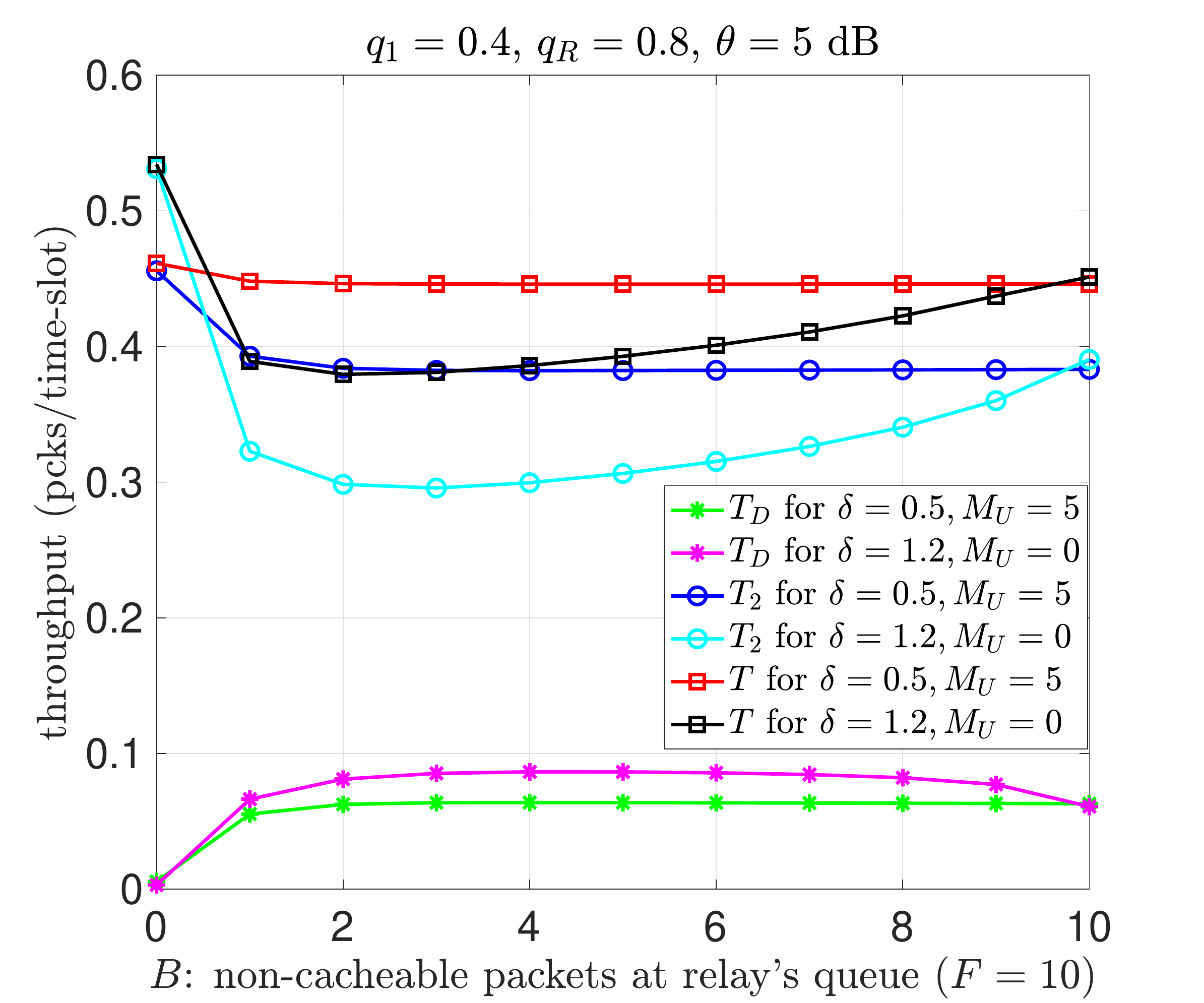}}\\
	\subfloat[$q_1 = 0.8$.]{
		\includegraphics[width=0.7\linewidth]{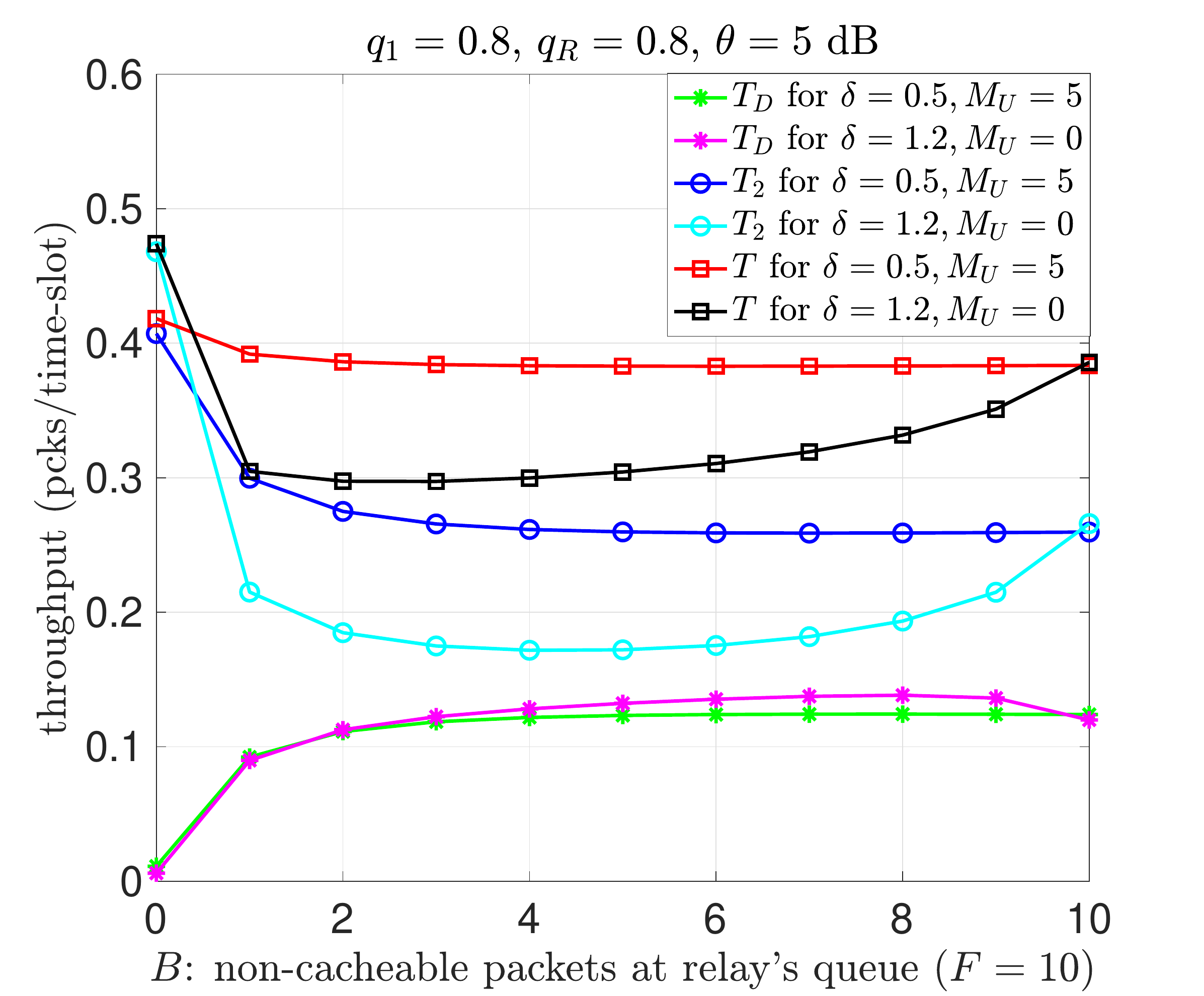}}
	\caption{Throughput $T_D, T_2,$ and $T$ vs. relay queue size $B$ (non-cacheable packets at the relay) when the relay node $R$ holds $ F = 10 $ files for both types of traffic, the links SINR $\theta = 5$ dB and either (i) $\delta = 0.5$ and $M_U=5$, or (ii) $\delta = 1.2$ and $M_U=0$.}
	\label{Fig:T_vs_B_theta_5}
\end{figure}

In this section, we study how $B$ i.e., the storage at the relay $R$ for non-cacheable traffic affects $T_D$, $T_2$, and $T$ i.e., the non-cacheable throughput at destination node $D$, the cacheable throughput at user node $U_2$, and the network throughput given by: $T= T_D + T_S$, respectively.
We present two cases in each plot: (i) $\delta = 0.5$ and $M_U = 5$, and (ii) $\delta = 1.2$ and $M_U = 0$. 
It should be noted that $q_U$ i.e., the probability of requesting content from external resources decreases with $\delta$ for fixed $M_U$, and $q_U=1$ for $M_U = 0$ (see Section \ref{Sec:CachePolicy}).
In Fig. \ref{Fig:T_vs_B_theta_5}(a) and (b), we plot the aforementioned throughputs versus $B$ for $q_1 = 0.4$ and $q_1 = 0.8$, respectively.
We observe that $T_D$ is increased with $q_1$ since increasing $q_1$ results in $U_1$ attempting transmissions more frequently. As a result, more interference to $U_2$ is realized from $U_1$ and $R$, and, hence, $T_2$, is decreased with $q_1$. Additionally, we observe that the network throughput $T$ is decreased with $q_1$.

Furthermore, when $q_1 = 0.4$, we observe that increasing $B$ above a minimum value is not beneficial for the three throughputs when $\delta = 0.5$ and $M_U = 5$. 
However, this is not the case when $\delta = 1.2$ and user $U_2$ has no cache i.e.,  $q_U = 1$. We observe that $T_2$ obtains its maximum value when $B=0$ i.e., when $R$ is not set to hold any non-cacheable packets, and, thus, is only available for serving cacheable traffic. There is a considerable drop in $T_2$, when $R$ is set to hold non-cacheable packets ($B>0$) as well, since $q_R$ i.e., the probability of $R$ being available to serve non-cacheable traffic, was set to $0.8$ in our results.  The network throughput $T$ behaves similarly to $T_2$.

When raising $q_1$ to $0.8$, we observe that $T_D$ increases with $B$ no matter the values for $\delta$ and $M_U$. 
Regarding $T_2$, we observe a similar behavior to the case in which $q_1=0.4$, but with lower values.
The slight decrease of $T_2$ with $B$ that is observed for $B<4$ can be attributed to the fact that when the queue is increased, the cache size at the relay is decreased and, hence, user $U_2$ requests more frequently content from $DC$ instead of $R$. 
Similarly to $q_1=0.4$, $T_2$ starts to drop with increased $B$ up to $3$ and, then, increases with $B$.


\subsection{Distribution of relay's queue states}
In this section we study $\pi(i)$ i.e., the probability of the queue $Q$ at the relay being in a specific state $i$ or, equivalently, holding $i$ non-cacheable packets, versus $i$ for the cases of Section \ref{Sec:Throughputs}: (i) $\delta = 0.5$ and $M_U = 5$, and (ii) $\delta = 1.2$ and $M_U = 0$.
We used three different values for $B$ i.e., the size of the queue at the relay, to gain insight into its effect on the distribution of the queue states.

\begin{figure}
	\centering
	\subfloat[$q_1 =0.4$.]{
			\includegraphics[width=0.65\linewidth]{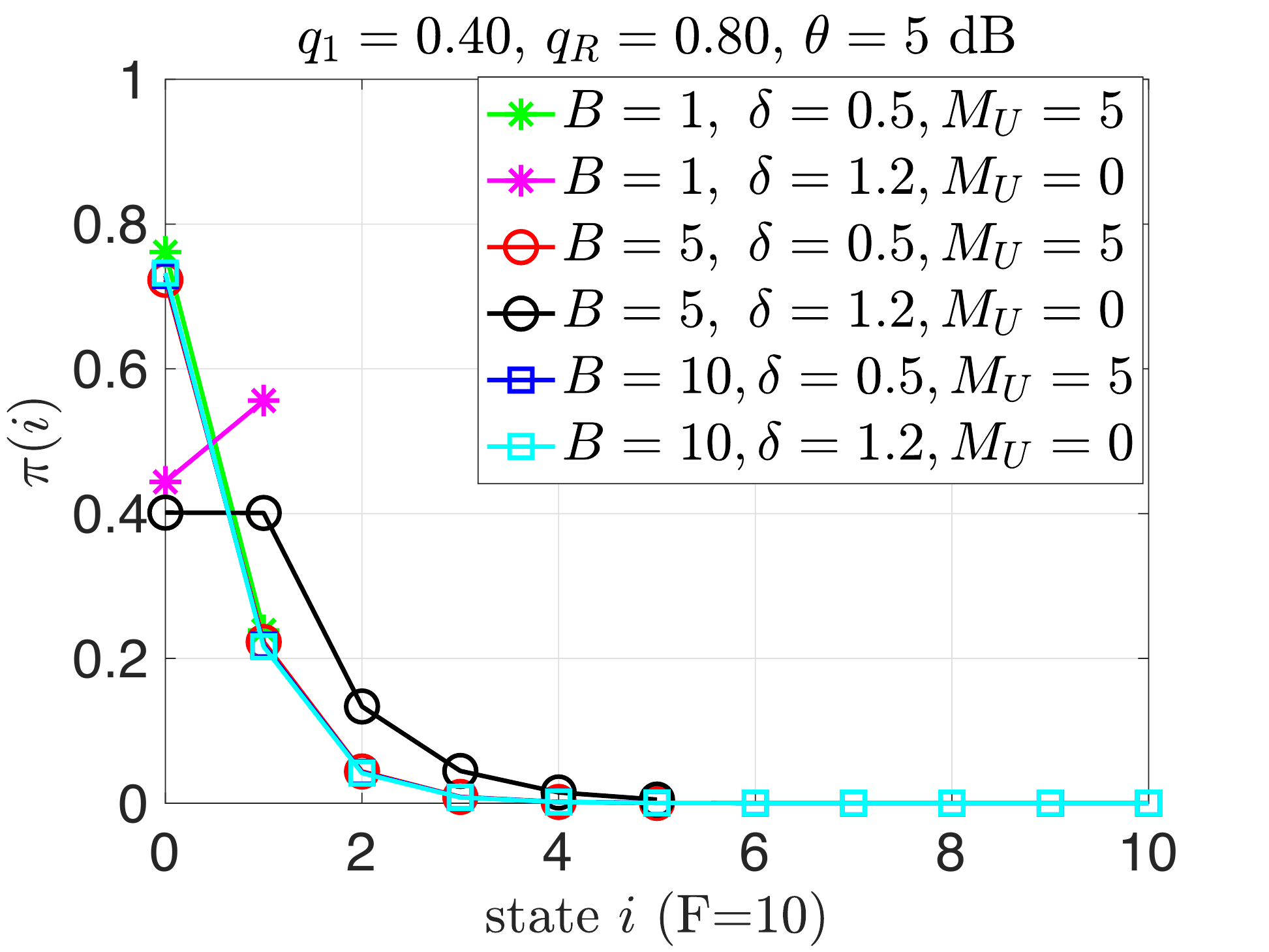}}\\
	\subfloat[$q_1 =0.8$.]{
		\includegraphics[width=0.65\linewidth]{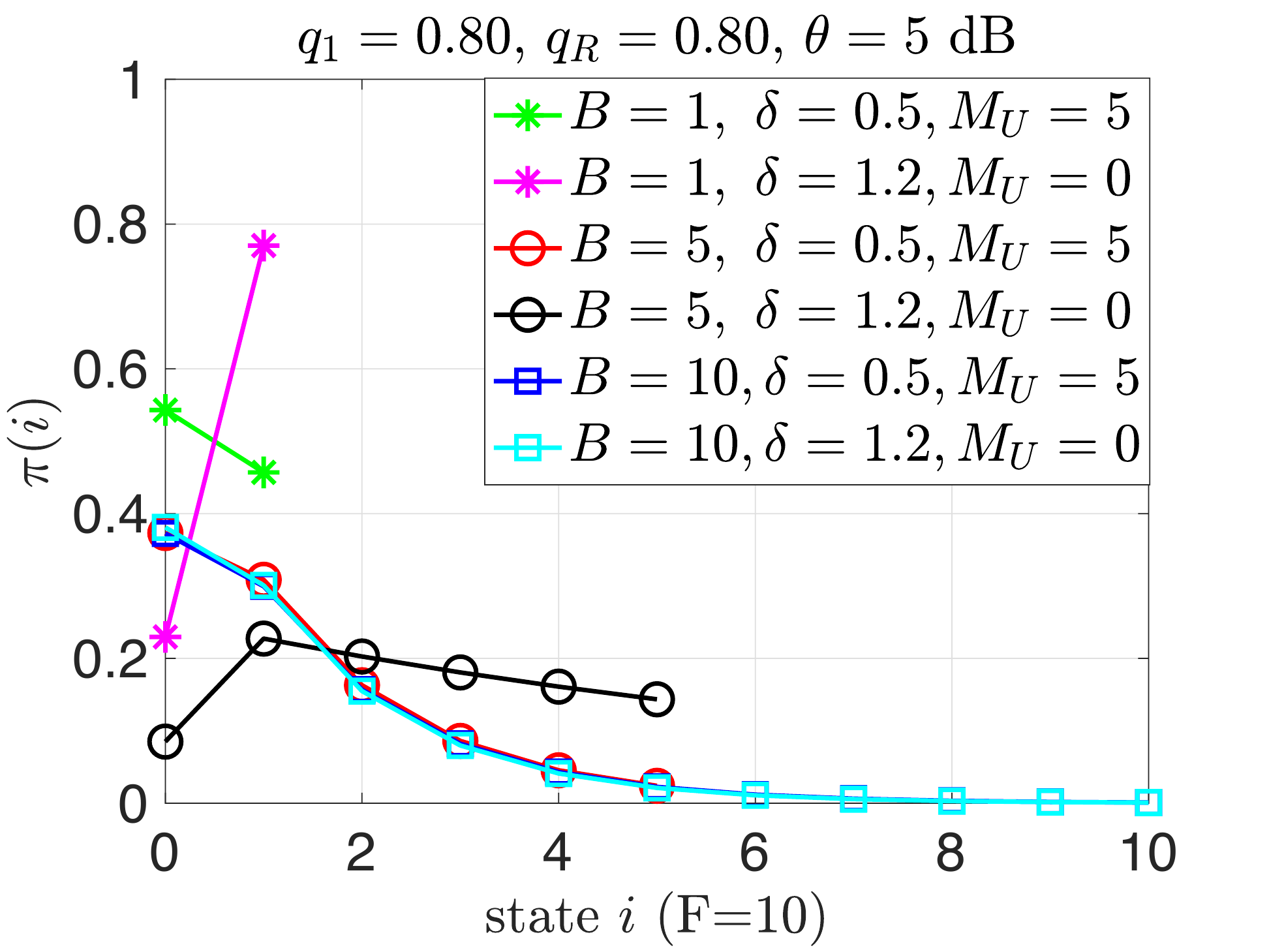}}
	\caption{Probability of the relay's queue $Q$ holding $i$ packets vs. $i$ when the relay node $R$ holds $ F = 10 $ files for both types of traffic, the links SINR $\theta = 5$ dB, and either (i) $\delta = 0.5$ and $M_U=5$, or (ii) $\delta = 1.2$ and $M_U=0$.}
	\label{Fig:Queue_theta_5}
\end{figure}

For $B=1$, it is more probable for the relay to store no packets at all when $\delta = 0.5$ and $M_U = 5$ and more probable to store one packet when $\delta = 1.2$ and $M_U = 0$ (see Fig. \ref{Fig:Queue_theta_5}).
Furthermore, when $B \in \{ 5, 10\}$, the queue is more probable to hold on average more packets when $\delta = 1.2$ and $M_U = 0$ than when $\delta = 0.5$ and $M_U = 5$ no matter the value of $q_1$.
We observe that, when $B =5$ i.e., the storage at the relay is equally split among cacheable and non-cacheable purposes, or when $B=10$ i.e., the storage at the relay is dedicated to non-cacheable traffic, then the probability of the queue holding more than $4$ packets is almost zero for $q_1 = 0.4$ (see Fig. \ref{Fig:Queue_theta_5}(a)). 
This is anticipated since, in our results, $q_R =0.8$ i.e., the probability by which the relay is available to serve non-cacheable traffic is double the rate by which non-cacheable traffic is transmitted to the network by $U_1$.

However, this is not the case if $q_1$ is increased to $0.8$.
In general, increasing $q_1$ yields higher values of $\pi(i)$ for higher states since increasing the rate by which $U_1$ attempts transmissions results in more failed transmissions to $D$ and, hence, more attempts to queue the failed packets at the relay. Consequently, the queue is more probable to store more packets for higher values of $q_1$ compared to lower ones.
Moreover, for fixed $q_1$, when $\delta=1.2$ and $M_U= 0$, increasing $B$ over five has a decreasing effect on $\pi(i)$ for higher states i.e., increasing $B$ produces a queue that has a higher probability of holding less packets.
This is not the case when $\delta = 0.5$ and $M_U = 5$ where the distributions of the queue states for the first six states are very close for $B \in \{5, 10\}$.

\section{Conclusion}
In this work, we studied the effect of a relay node with storage capabilities in a wireless system that serves two types of traffic: cacheable and non-cacheable traffic. 
The relay's storage can be split to accommodate the needs of both types of traffic.
We derived the network throughput taking into consideration the number of files for cacheable and non-cacheable traffic at the relay, the wireless links' parameters, the availability of the data center, and the rate by which cacheable and non-cacheable content is requested and transmitted, respectively.

Our numerical results provide insight into the network throughput and distribution of the relay's files in terms of the aforementioned parameters. 
It is shown how the network throughput is affected by allocation of the relay storage to cacheable and non-cacheable files as well as how the distribution of the cached files affects the network's performance. 

\bibliographystyle{IEEEtran}
\bibliography{ICC_camera_ready}{\tiny}

\end{document}